\documentclass[twocolumn,showpacs,amsmath,nofootinbib,pra,aps,amssymb,longbibliography]{revtex4-2}

\usepackage[T1]{fontenc}
\usepackage[utf8]{inputenc}
\usepackage{times}
\usepackage{color} 
\usepackage{array}
\usepackage{amssymb,amsmath}
\usepackage{amsbsy}
\usepackage[pdftex]{graphicx} 
\usepackage{bm}
\usepackage{float}
\usepackage{dcolumn}
\usepackage{scalerel}
\usepackage{tikz}
\usetikzlibrary{svg.path}

\definecolor{orcidlogocol}{HTML}{A6CE39}
\tikzset{
  orcidlogo/.pic={
    \fill[orcidlogocol] svg{M256,128c0,70.7-57.3,128-128,128C57.3,256,0,198.7,0,128C0,57.3,57.3,0,128,0C198.7,0,256,57.3,256,128z};
    \fill[white] svg{M86.3,186.2H70.9V79.1h15.4v48.4V186.2z}
                 svg{M108.9,79.1h41.6c39.6,0,57,28.3,57,53.6c0,27.5-21.5,53.6-56.8,53.6h-41.8V79.1z M124.3,172.4h24.5c34.9,0,42.9-26.5,42.9-39.7c0-21.5-13.7-39.7-43.7-39.7h-23.7V172.4z}
                 svg{M88.7,56.8c0,5.5-4.5,10.1-10.1,10.1c-5.6,0-10.1-4.6-10.1-10.1c0-5.6,4.5-10.1,10.1-10.1C84.2,46.7,88.7,51.3,88.7,56.8z};
  }
}

\newcommand\orcidicon[1]{\href{https://orcid.org/#1}{\mbox{\scalerel*{
\begin{tikzpicture}[yscale=-1,transform shape]
\pic{orcidlogo};
\end{tikzpicture}
}{|}}}}

\usepackage[unicode,breaklinks]{hyperref}
\hypersetup{
    unicode=true,
    plainpages=false, 
    colorlinks=true,
    linkcolor=blue,
    citecolor=blue,
    filecolor=black,
    urlcolor=blue
}
\usepackage{url}
\usepackage{verbatim}

\newcommand{\add}{a_\mathrm{dd}}
\newcommand{\edd}{\epsilon_\mathrm{dd}}
\newcommand{\br}{\mathbf{r}}
\newcommand{\bx}{\mathbf{x}}

\newcommand\gammaQF{\gamma_\mathrm{QF}}

\begin{document}
 
\title{Sounds waves and fluctuations in one-dimensional supersolids} 
\author{L.~M.~Platt\orcidicon{0009-0002-2590-3153}}
\author{D.~Baillie\orcidicon{0000-0002-8194-7612}}
\author{P.~B.~Blakie\orcidicon{0000-0003-4772-6514}} 
	\affiliation{%
	$^1$Dodd-Walls Centre for Photonic and Quantum Technologies, Dunedin 9054, New Zealand\\
	$^2$Department of Physics, University of Otago, Dunedin 9016, New Zealand}

\date{\today} 
\begin{abstract}  
We examine the low-energy excitations of a dilute supersolid state of matter with a one-dimensional crystal structure. A hydrodynamic description is developed based on a Lagrangian, incorporating generalized elastic parameters derived from ground state calculations. The predictions of the hydrodynamic theory are validated against solutions of the Bogoliubov-de Gennes equations, by comparing the speeds of sound, density fluctuations, and phase fluctuations of the two gapless bands. Our results are presented for two distinct supersolid models: a dipolar Bose-Einstein condensate in an infinite tube and a dilute Bose gas of atoms with soft-core interactions. Characteristic energy scales are identified, highlighting that these two models approximately realize the bulk incompressible and rigid lattice supersolid limits. 
\end{abstract} 

\maketitle
\section{Introduction}
A supersolid is a state of matter in which superfluid and crystalline properties coexist as a consequence of the simultaneous breaking of phase and translational symmetries. The properties of supersolids have been discussed in theoretical works for almost 50 years, prior to their realizations in ultra-cold atomic systems \cite{Leonard2017a,Li2017a,Tanzi2019a,Bottcher2019a,Natale2019a}. Experiments with these supersolids have investigated their basic phase diagram, excitations and dynamics. The excitation spectrum is of great interest, because the broken translational symmetry results in new gapless excitation branches 
called Nambu-Goldstone modes \cite{Watanabe2012a}.  
For the case of supersolids with one-dimensional (1D) crystalline structure, two gapless excitation bands have been found. 
These branches can be classified by the character of fluctuations they cause  \cite{Macri2013a,Ancilotto2013a,Natale2019a}, with the lower energy branch being associated with phase fluctuations and particle tunneling between lattice sites, and the upper branch being more strongly associated with density fluctuations arising from the crystal motion \cite{Roccuzzo2019a,Tanzi2019b,Guo2019a,Natale2019a,Petter2021a,Ilg2023a}.  
Quantitative descriptions of the excitations involve large-scale numerical calculations and do not yield much insight into the underlying physics.  However, hydrodynamic theories offer the possibility to understand the low-energy and long-wavelength aspects of the supersolids, e.g.~the speeds of sounds and nature of the  fluctuations for each branch, in terms of fundamental properties of the system. Work on such theories for  supersolids dates back to the seminal 1969 paper by Andreev and Lifshitz \cite{Andreev1969a}, and extended by Saslow \cite{Salsow1977a} and Liu \cite{Liu1978a}. More recently there has been work on deriving an effective Lagrangian for supersolids by various means. Son \cite{Son2005a} has utilized the invariances and broken symmetries of the supersolid state to provide a universal description also see \cite{Yoo2010a,Hofmann2021a} and \cite{Buhler2023a}. Josserand \textit{et al.}~\cite{Josserand2007a,Josserand2007b} have obtained an effective Lagrangian by applying the technique of homogenisation to a nonlocal Gross-Pitaevskii theory that predicts crystallization (also see  Refs.~\cite{Ye2008a}  and \cite{Peletminskii2008a}).  In recent work \v{S}indik \textit{et al.}~\cite{Sindik2023a} have presented a practical proposal for experiments to investigate the hydrodynamic properties and the superfluid fraction using a dipolar supersolid in a toroidal trap by using a long-wavelength density-coupled perturbation.  We also note a recent experiment that has used an optical lattice to excite a self-induced Josephson effect and quantify the superfluid fraction in a dipolar supersolid \cite{Biagioni2023a}.

\begin{figure}[htbp]
	\centering
	\includegraphics[width=3.4in]{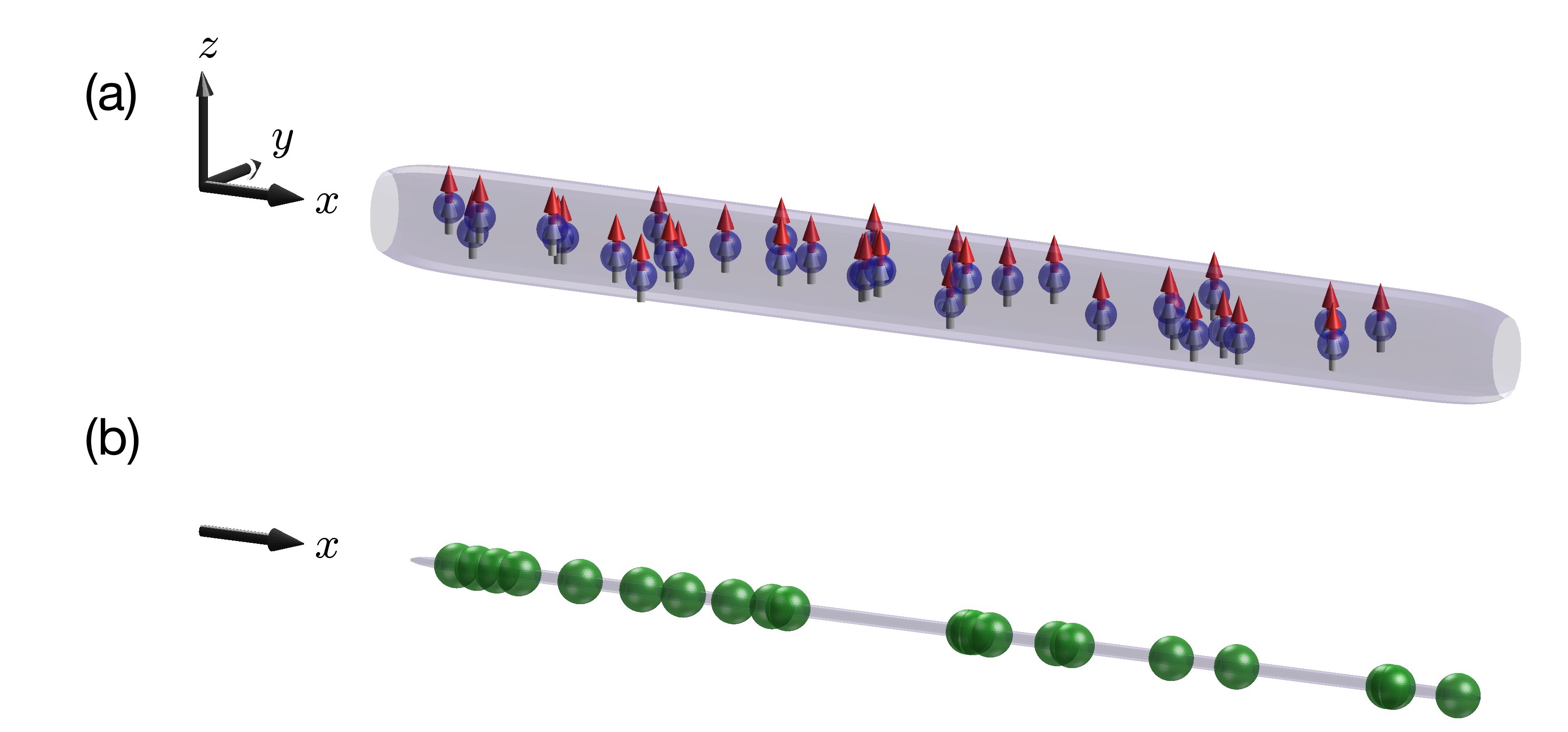}
	\caption{Schematic of systems we consider: (a) A dipolar BEC confined in an infinite tube potential with magnetic dipole moments polarized along $z$, and  (b) a BEC of particles with soft-core interactions. Both systems have unconfined motion along $x$ and in the supersolid transition translational invariance is broken along this axis.
	\label{figschematic}}
\end{figure}

Main motivation of this work is to provide the first quantitative test of a hydrodynamic description of a supersolids by comparing its predictions to the direct numerical calculation of the excitations for specific models. The hydrodynamic theory depends a set of generalized elastic parameters including the superfluid fraction, compressibility and lattice elastic parameters, and we show how to determine these from ground state calculations. 
 In order to illustrate our results we consider two contrasting systems [see Fig.~\ref{figschematic}]. First, a dipolar Bose-Einstein condensate (BEC) in a tube shaped potential where transition to a 1D supersolid occurs as the $s$-wave scattering length is changed. This system is the thermodynamic limit corresponding to experiments that have produced  supersolids in cigar shaped potentials. Second, a 1D soft-core Bose gas, which has been studied as a basic model of supersolidity \cite{Sepulveda2008a,Kunimi2011a,Prestipino2019a} and its properties contrast those of the dipolar case. There are proposals for producing a BEC with soft-core interactions using  atoms that are weakly coupled to a highly excited Rydberg state \cite{Henkel2010a}.  
 Our results for these two models show that they have markedly different properties. The dipolar supersolid arises from a mechanical instability with the condensate partially collapsing to higher density incompressible state. In contrast, the purely repulsive interactions of the soft-core model the supersolid arising from a clustering transition, leading to a relatively rigid lattice.  These two cases are conveniently distinguished by comparing the energy scales associated with the isothermal compressibility at constant strain and the longitudinal elastic modulus of the lattice.

 The outline of the paper is as follows. In Sec.~\ref{Sec:Systems} we briefly introduce the dipolar and soft-core systems we use in this work. The supersolid Lagrangian is introduced in Sec.~\ref{Sec:Lagrangian} and we discuss the how to obtained the elastic parameters that appear in the Lagrangian. In Sec.~\ref{Sec:Excitations} we derive the hydrodynamic behavior of the excitations from the Lagrangian and use this to determine the speeds of sound. We then make a detailed comparison to calculations of the two models, and explore the limiting behavior of the predictions. The density fluctuations are discussed in Sec.~\ref{Sec:DenFlucts}. Here a focus is on the decomposition of the fluctuations into components related to tunneling of atoms between sites (defect density fluctuations) and due movement of the crystal lattice (lattice strain fluctuations). Finally, in Sec.~\ref{Sec:PhaseFlucts} we discuss the phase fluctuations before concluding in Sec.~\ref{Sec:Conclusions}.

\section{Systems}\label{Sec:Systems}
We consider two zero-temperature systems that undergo a transition to a 1D crystalline state.   The models are briefly introduced in the following two  subsections. Additional details of the models and numerical solution method are given in Appendix \ref{App:SSmodels}. Apart from the microscopic parameters particular to each model, we consider both systems in a thermodynamic limit with a specified density per unit length $\rho$.  

\subsection{Dipolar Bose gas in a tube potential}
Our first system is a  Bose gas of magnetic atoms confined in a tube shaped potential  [see  Fig.~\ref{figschematic}(a)]. The single particle Hamiltonian is
\begin{equation}
	H_\mathrm{sp} = -\frac{\hbar^2\nabla^2}{2m} + \frac12 m\omega^2(y^2+z^2),
\end{equation}
where $\omega$ is the transverse harmonic confinement angular frequency, and the atoms are free to move along the $x$-axis.
The atomic dipole moments are polarized along $z$ by a bias field and the interactions are described by the potential 
\begin{equation}
	U(\br) = \frac{4\pi a_s\hbar^2}{m}\delta(\br) + \frac{3\add\hbar^2}{m r^3}\left(1-3\frac{z^2}{r^2}\right),
\end{equation}
where $a_s$ is the $s$-wave scattering length,  $\add = m\mu_0\mu_m^2/12\pi\hbar^2$ is the dipole length, and $\mu_m$  is the atomic magnetic moment. 
 The ratio $\edd=\add/a_s$ characterises the relative strength of the dipole-dipole to s-wave interactions. When this parameter is sufficiently large the ground state undergoes a transition to a crystalline state with modulation along $x$. Quantum fluctuation effects are necessary to stabilize this state and further details about this model are given in Appendix \ref{App:DEPGE}. The value of $\edd$ where the transition occurs depends on $\omega$ and $\rho$, and the transition can be continuous or discontinuous. In the results we present here are for $^{164}$Dy atoms with  $\add=130.8\,a_0$ and  $\omega/2\pi=150\,$Hz.

\subsection{1D soft-core Bose gas}
Our second system is a BEC of atoms free to move in 1D, but interacting via a soft-core interaction potential  $U_{\mathrm{sc}}(x)=U_0\theta(a_\mathrm{sc}-|x|)$, where $a_\mathrm{sc}$ is the core radius and $U_0$ is the potential strength  [see  Fig.~\ref{figschematic}(b)].  It conventional to define the dimensionless interaction parameter
\begin{align}
\Lambda = \frac{2ma_{\mathrm{sc}}^3U_0\rho}{\hbar^2},
\end{align}
and continuous transition occurs from a uniform to a modulated state at the critical value $\Lambda_c=21.05$.  Further details about this model are given in Appendix \ref{App:SC}.
 
\subsection{General solution properties}
By solving the (generalized) Gross-Pitaevskii equations for these models we determine the energy minimising wavefunction $\psi_0$. These solutions can be found in a single unit cell of length $a$ along $x$. If the state is crystalline, then $a$ is the lattice constant (for uniform states the energy is independent of $a$). Take $\mathcal{E}$ to be the energy per unit length of this state for density $\rho$ and lattice constant $a$ (as defined in Appendix  \ref{App:SSmodels}). In the thermodynamic limit where the system is free to choose the microscopic length $a$ to minimise the energy, the ground state is
 \begin{align}
 \mathcal{E}_0(\rho)=\min_{a} \mathcal{E}(\rho,a),\label{E_0}
 \end{align}
 which can be used to implicitly determine $a(\rho)$ as the value of $a$ that minimises $\mathcal{E}$ at density $\rho$.  
 
\section{Supersolid Lagrangian}\label{Sec:Lagrangian}
\subsection{Quadratic Lagrangian density}
Here our interest is in a Lagrangian description of the low-energy, long-wavelength features of a system where 1D crystalline order can occur along the $x$-direction. We work in a thermodynamic limit where the ground state is determined by the average linear number density $\rho$ in addition to the interaction and relevant transverse confinement parameters. Long-wavelength perturbations of the ground state can be characterized by three fields:  $\{\delta \rho(x,t),u(x,t), \phi(x,t)\}$. Here $\delta \rho(x,t)$ denotes variations in the average density, $u$ denotes the deformation field of the crystal lattice along the $x$ direction (relative to the unstrained ground state), and $\phi$ denotes the superfluid phase field.  For treating small departures from equilibrium the Lagrangian density can be expanded to quadratic order in these fields as 
\begin{align}
\mathcal{L}=&-\hbar\,\delta \rho\,\partial_{t}\phi-\frac{\alpha_{\rho\rho}}{2}(\delta \rho)^{2}-\frac{\alpha_{uu}}{2}(\partial_{x}u)^{2}-\alpha_{\rho u}\delta \rho\,\partial_{x}u \nonumber \\ 
&+\frac{1}{2}m\rho_{n}\Bigl(\partial_{t}u-\frac{\hbar}{m}\partial_{x}\phi\Bigr)^{2}-\rho\frac{\hbar^{2}}{2m}(\partial_{x}\phi)^{2},\label{LagrangianDensity}
\end{align}
where $\rho_n$ is the normal density, with $\rho=\rho_s+\rho_n$, and $\rho_s$ being the superfluid density. Similar forms of Lagrangian densities have been obtained by other authors  (e.g.~see \cite{Josserand2007a,Josserand2007b,Yoo2010a,Hofmann2021a,Buhler2023a}), although here we follow the approach of Yoo and Dorsey \cite{Yoo2010a}. A discussion about the relationship between these treatments is given in Refs.~\cite{Yoo2010a,Hofmann2021a}.
The phase gradient relates to the superfluid velocity as $v_s=\frac{\hbar}{m}\partial_x\phi$ and the normal velocity relates to the time-derivative of the lattice deformation field $v_n=\partial_tu$ \cite{Son2005a,Yoo2010a}.

\subsection{Generalized elastic parameters}\label{Sec:GElastic}
 
In addition to the superfluid density, which determines the rigidity of the state to twists in the phase $\phi$ \cite{Fisher1973a}, three other generalized elastic parameters of the system $\{\alpha_{\rho\rho},\alpha_{\rho u},\alpha_{uu}\}$ also appear in Eq.~(\ref{LagrangianDensity}). Here we discuss how these can  be obtained from ground state calculations.

For 1D crystals the superfluid fraction $f_s=\rho_s/\rho$ has  \begin{align}
 f_s^+&=\frac{a}{\rho}\left[\int_\mathrm{uc}\frac{dx}{\int dy\,dz\,|\psi_0|^2}\right]^{-1},\\
 f_s^-&=\frac{a}{\rho}\int dy\,dz\left[\int_\mathrm{uc}\frac{dx}{|\psi_0|^2}\right]^{-1},
 \end{align}
 as upper and lower bounds, respectively \cite{Leggett1970a,Leggett1998a}. Hereon, we neglect the transverse coordinates $(y,z)$ when applying results to the soft-core case.
 Both bounds are identical for the soft-core case, and provide the exact superfluid fraction (e.g.~see \cite{Sepulveda2008a}). For the dipolar case the bounds are close to each other and taking $f_s=\frac{1}{2}( f_s^+ + f_s^-)$ provides an accurate estimate of the superfluid fraction (see \cite{Smith2023a}).

The other generalized elastic parameters can be determined from the energy density $\mathcal{E}(\rho,a)$, for the ground state under the constraint of lattice constant being $a$ and the mean density being $\rho$. 
The first elastic parameter is defined by
\begin{align}
\alpha_{\rho\rho}=\left(\frac{\partial ^2\mathcal{E}}{\partial \rho^2}\right)_a,\label{alpharr}
\end{align}
and relates to the isothermal compressibility at constant strain:  
\begin{align}
\tilde\kappa=\frac{1}{\rho^2\alpha_{\rho\rho}}.\label{kappatilde}
\end{align}
The second parameter describes the effects of straining a site at constant density
\begin{align}
\alpha_{uu}=a^2\left(\frac{\partial ^2\mathcal{E}}{\partial a^2}\right)_\rho.\label{alphauu}
\end{align}
Noting that lattice strain, given by $\partial_xu$, relates to changes in the lattice constant according to $\delta a/a=\partial_xu$. 
 In higher dimensional crystals the $\alpha_{uu}$ term generalizes to the elastic tensor. For the 1D crystal   $\alpha_{uu}$ can also be identified as the layer-compressibility used to characterize smectic materials \cite{Chaikin1995a,Hofmann2021a}.
Finally,  we  define the density-strain coupling parameter by the mixed partial derivative
\begin{align}
\alpha_{\rho u}=a\left(\frac{\partial ^2\mathcal{E}}{\partial \rho \partial a}\right).\label{alpharu}
\end{align}

\section{Excitations and speeds of sound}\label{Sec:Excitations}
 In this section we consider the collective excitations of the gapless energy bands. We begin by deriving hydrodynamic results for the collective modes, and the associated speeds of sound. We then compare these to results for our two models, and consider the limiting behavior.  
 
\subsection{Hydrodynamic equations of motion and collective modes}
 The Euler-Lagrange equations obtained from Eq.~(\ref{LagrangianDensity}) are
\begin{align}
   \hbar\partial_t\phi &=- \alpha_{\rho\rho}\delta\rho - \alpha_{\rho u} \partial_x u, \label{ELdrho}\\
   \rho_n(m\partial^2_tu-\hbar\partial_{tx}\phi) &= \alpha_{uu}\partial^2_xu+\alpha_{\rho u} \partial_x \delta\rho,  \\
   \partial_t\delta\rho   &=    -\rho_n\partial_{tx}u-\rho_s\frac{\hbar}m\partial^2_x\phi. \label{ELu}
\end{align}
These describe the hydrodynamic evolution for small departures from equilibrium.

We look for normal mode solutions of Eqs.~(\ref{ELdrho}) - (\ref{ELu}) of the form $X(x,t) =X_w e^{i(qx-\omega t)}$, where $X$ denotes our three fields appearing in the Lagrangian and $q$ is a quasimomentum. The resulting linear system is
\begin{align} 
    \mathsf{M}
    \begin{pmatrix}
        \delta\rho_w \\ \phi_w \\ u_w
    \end{pmatrix}
=\mathbf0,
\end{align}
where 
\begin{align}
  \mathsf{M} &= \begin{pmatrix}
 \alpha_{\rho\rho} & -i\hbar\omega  & iq\alpha_{\rho u}  \\
i\hbar\omega  &   \hbar^2 q^2 \rho_s/m  & -\hbar\omega q \rho_n  \\
iq \alpha_{\rho u} & \hbar\omega q\rho_n   &\omega^2\rho_nm -q^2\alpha_{uu}
    \end{pmatrix}.
 \end{align}
Nontrivial solutions require the matrix $\mathsf{M}$ to be singular, i.e.~its determinant $\Delta_\mathsf{M}$ must be zero, where 
\begin{align}
\Delta_\mathsf{M}=m\hbar \rho_n(\omega^2-c_+^2q^2)(\omega^2-c_-^2q^2).
\end{align}
Here we have introduced 
\begin{align}
mc_{\pm}^{2}&=\frac{1}{2}(a_\Delta\pm\sqrt{a_\Delta^2-4b_\Delta}),\label{Eqcpm}
\end{align}
where
\begin{align}
a_\Delta&=\rho\alpha_{\rho\rho}-2\alpha_{\rho u}+\frac{\alpha_{uu}}{\rho_n},\\
b_\Delta&=\frac{\rho_s}{\rho_n}(\alpha_{\rho\rho}\alpha_{uu}- \alpha_{\rho u}^2).
\end{align}
Thus the hydrodynamic excitation frequencies of the two solutions are gapless and given by
\begin{align}
\omega_{\pm}=c_\pm q,
\end{align}
with  $c_\pm$ being the speeds of sound.

\subsection{Results for the speeds of sound}
 \begin{figure}[htbp]
	\centering
	\includegraphics[width=3.3in]{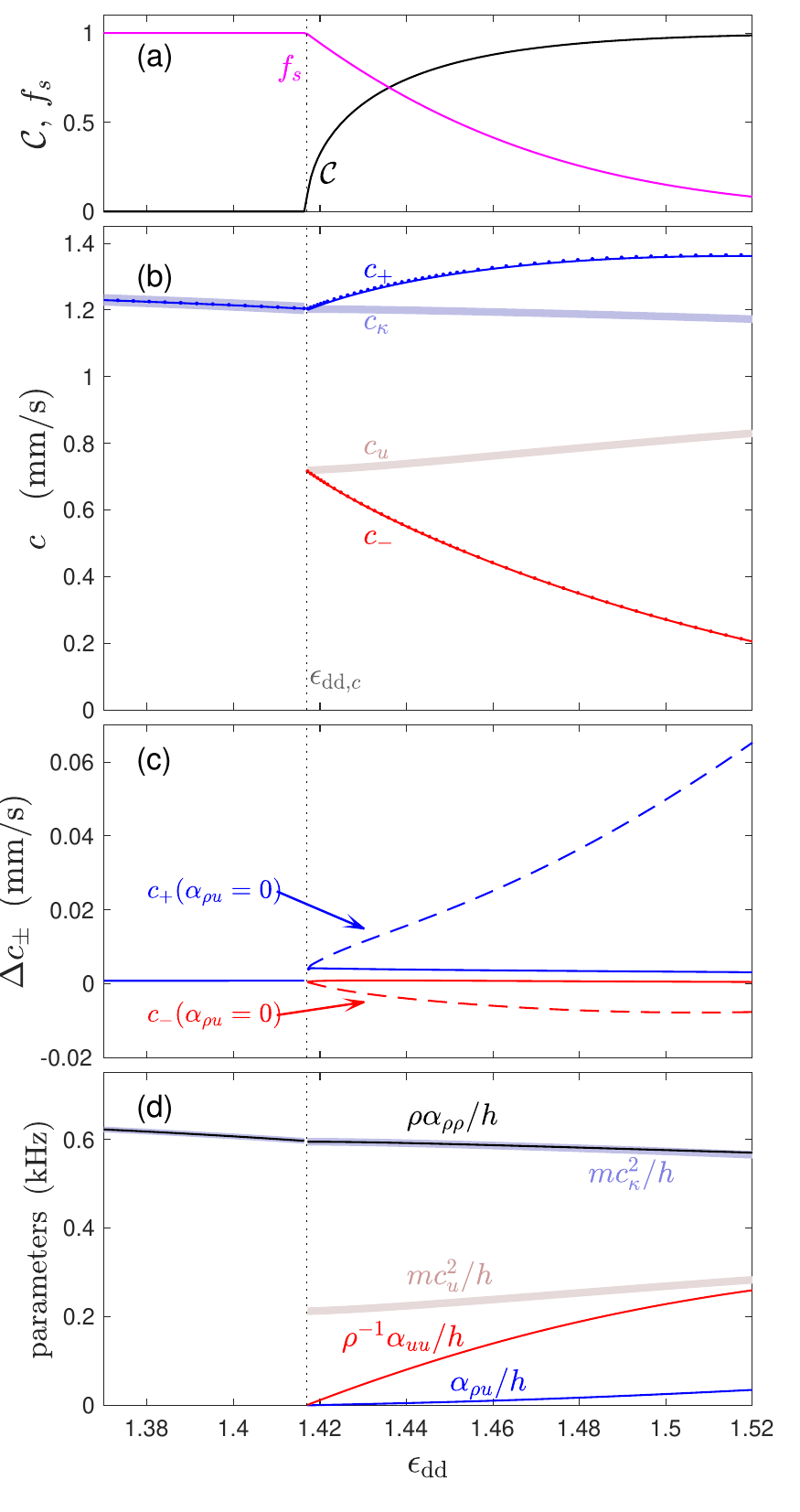}
	\caption{Speeds of sound   for a dipolar gas.
	(a) Density contrast and superfluid fraction.  (b) Speeds of sound across the superfluid to supersolid transition. BdG results  (solid lines) and analytic result (\ref{Eqcpm}) similarly colored dots. For reference 
	$c_\kappa$ (thick light blue line) and $c_{u}$ (thick light red line) are shown. (c) The difference of Eq.~(\ref{Eqcpm}) from the BdG speeds of sound (solid lines) using same colors as in (a). Also shown is this difference setting $\alpha_{\rho u}=0$ (dashed lines) in Eq.~(\ref{Eqcpm}). (d) The generalized elastic parameters compared to $mc_\kappa^2$ and $mc_{u}^2$. Other parameters: $^{164}$Dy atoms with $\rho=2500/\mu$m and radially symmetric transverse confinement of $\omega_{y,z}/2\pi=150\,$Hz.
	\label{figDipsound}}
\end{figure}

\begin{figure}[htbp]
	\centering
	\includegraphics[width=3.3in]{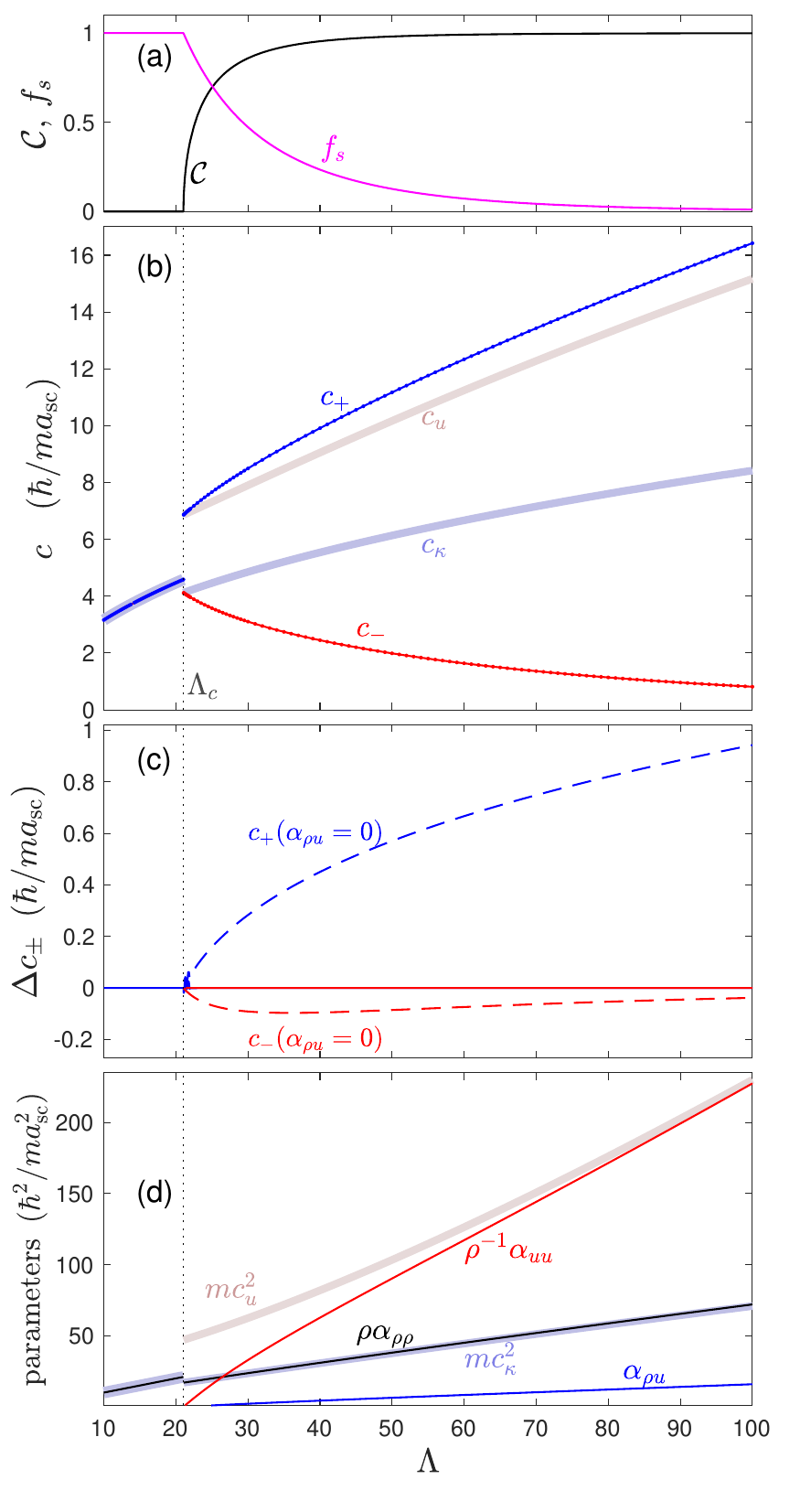}
	\caption{Speeds of sound for a soft-core gas.
	(a) Density contrast and superfluid fraction. (b) Speeds of sound across the superfluid to supersolid transition. BdG results  (solid lines) and analytic result (\ref{Eqcpm}) similarly colored dots. For reference 
	$c_\kappa$ (thick light blue line) and $c_{u}$ (thick light red line) are shown. (c) The difference of Eq.~(\ref{Eqcpm}) from the BdG speeds of sound (solid lines) using same colors as in (a). Also shown is this difference setting $\alpha_{\rho u}=0$ (dashed lines) in Eq.~(\ref{Eqcpm}). (d) The generalized elastic parameters compared to $mc_\kappa^2$ and $mc_{u}^2$.	\label{figSCsound}}
\end{figure}

We compare the speeds of sound determined from direct calculation of the excitations to the hydrodynamic result of Eq.~(\ref{Eqcpm}) in Figs.~\ref{figDipsound} and \ref{figSCsound} for  the dipolar and soft-core systems, respectively. The direct calculation of the excitations involves solving the Bogoliubov-de Gennes (BdG) equations (e.g.~see \cite{Kunimi2012a,Macri2013a,Ancilotto2013a,Roccuzzo2019a,Blakie2023a}).
We make this comparison across the uniform superfluid to supersolid transition.
In Figs.~\ref{figDipsound}(a) and \ref{figSCsound}(a) we show the density contrast $\mathcal{C}$ defined as
\begin{align}
\mathcal{C}=\frac{\max\varrho_0-\min\varrho_0}{\max\varrho_0+\min\varrho_0},
\end{align}
where $\varrho_0(x)=\int dydz|\psi_0(\mathbf{x})|^2$ is the line density along $x$. The contrast  reveals the appearance of density modulations, and hence acts as an order parameter for the formation of crystalline order. In the dipolar system, at the densities we consider here \cite{Blakie2020b,Blakie2023a}, the crystalline order develops continuously as $\epsilon_{\mathrm{dd}}$ increases past the critical value of $\epsilon_{\mathrm{dd},c}$. For the soft-core system the transition is also continuous, and occurs when the interaction parameter $\Lambda$ exceeds $\Lambda_c$. In both systems, the superfluid fraction decreases with increasing density contrast.
 
The numerical  solution of the BdG equations \cite{Blakie2023a} yields the energies $\hbar\omega_{\nu q}$ and amplitude functions $\{\mathrm{u}_{\nu q},\mathrm{v}_{\nu q}\}$, where $\nu$ is the band index and $q$ is the quasimomentum along $x$. We then obtain the speeds of sound $c_\nu$ of gapless branches from the BdG results by fitting 
$\omega_{\nu q}$ to $c_\nu q$, 
 for small values of $q$. In the uniform superfluid phase only a single   branch is gapless, and we identify the speed of sound of this branch as  $\nu=+$. In the crystalline state there are two gapless branches, and we assign the $\nu$ of the lower (upper) branches as $-$   ($+$). Our results for the speeds of sound are shown in Figs.~\ref{figDipsound}(b) and \ref{figSCsound}(b) for the dipolar and soft-core systems, respectively. There are significant differences in behavior revealed by these results, particularly in the nature of the discontinuity of the speeds of sound at the transition point. For the dipolar system $c_+$ jumps downwards as $\epsilon_{\mathrm{dd}}$ crosses the transition\footnote{For the $\rho$ considered the jump is small and hard to see. Ref.~\cite{Blakie2023a} shows results at other densities where the discontinuity is more clearly revealed.}. In contrast, for the soft-core system the  $c_+$ speed of sound jumps upwards at the transition.
Other aspects  of the behavior near the transition will be discussed further in Sec.~\ref{Sec:limitbehave}. 

Since the BdG speeds of sound and the hydrodynamic  results are in good agreement,  in  Figs.~\ref{figDipsound}(c) and \ref{figSCsound}(c) we show the difference between the   results. The relative error is $\lesssim 1\%$ for the dipolar gas and significantly smaller for the soft-core model, except very close to the transition. The error for the dipolar model is close to the accuracy that we can determine the speeds of sound from the BdG calculations.  We also show the difference arising from neglecting the density-strain term (i.e.~setting $\alpha_{\rho u}=0$, which makes our theory equivalent to that in Refs.~\cite{Hofmann2021a,Sindik2023a}), noting that over the parameter range considered this causes a $\sim5\%$ relative upward-shift in the $c_+$ speed of sound and a smaller change in $c_-$. Furthermore, the role of the density-strain term vanishes as we approach the transition.

The hydrodynamic results presented here are determined by generalized elastic parameters (i.e.~$\{f_s,\alpha_{\rho\rho},\alpha_{\rho u},\alpha_{uu}\}$), which are determined from the ground state calculations as discussed in Sec.~\ref{Sec:GElastic}. The superfluid fraction results are shown in Figs.~\ref{figDipsound}(a) and \ref{figSCsound}(a), and the other elastic parameters are shown in Figs.~\ref{figDipsound}(d) and \ref{figSCsound}(d). These results reveal that the   crystal-dependent elastic terms $\alpha_{\rho u}$ and $\alpha_{u u}$ both vanish as we approach the transition from the modulated side.  For both systems we find that $\rho\alpha_{\rho u}$ is typically smaller than $\alpha_{u u}$, suggesting that neglecting this parameter can be a reasonable first approximation.
The  parameter $\alpha_{\rho\rho}$ is non-zero in both phases, but changes discontinuously at the transition (also see \cite{Blakie2023a}).

\subsection{Supersolid sound limiting behavior}\label{Sec:limitbehave}
From the hydrodynamic results for the speed of sound in Eq.~(\ref{Eqcpm}) we derive limiting results to describe their behavior approaching the transition and deep into the crystal regime. It is useful to introduce two energy scales   as
 \begin{align}
 mc_{\kappa}^2 &\equiv \rho \frac{ \alpha_{\rho\rho}\alpha_{uu}-\alpha_{\rho u}^2}{\alpha_{uu}}=\frac{1}{\rho\kappa}, \label{ckappa}\\
 mc_{u}^2 &\equiv \frac {\alpha_{uu}}{\rho_n}\label{cuu},
 \end{align}
being the bulk compressibility and lattice compressibility energies, respectively, defining the associated characteristic speeds $c_{\kappa}$ and $c_{u}$. Here 
 \begin{align}
 \kappa=\frac{\alpha_{uu}}{\rho^2(\alpha_{\rho\rho}\alpha_{uu}-\alpha_{\rho u}^2)},\label{kappa}
 \end{align}
is the isothermal compressibility. Because $\alpha_{\rho u}$ is relatively small in both our models $ \kappa\approx \tilde \kappa$ [see Eq.~(\ref{kappatilde}), Figs.~\ref{figDipsound}(d) and \ref{figSCsound}(d)]. The lattice compressibility is only non-zero in the modulated state. While both $\alpha_{uu}$ and $\rho_n$ go to zero as we approach the transition from the modulated side,  $mc_{u}^2$  has a non-zero value [see  Figs.~\ref{figDipsound}(d) and \ref{figSCsound}(d)].

 \subsubsection{Uniform superfluid}
In the uniform superfluid  $f_s=1$, $\alpha_{uu}=\alpha_{\rho u}=0$ and $\kappa\to1/\rho^2\alpha_{\rho\rho}=\tilde{\kappa}$. In this case Eq.~(\ref{Eqcpm}) yields the single nontrivial speed of sound  $c_+=c_\kappa$.  This situation is well-known for BECs, with the speed of sound being directly related to the isothermal compressibility  \cite{BECbook}. 
 
 \subsubsection{Approaching the transition}\label{SecApproachTransition}
 Approaching the transition from the modulated side, the density contrast vanishes and the superfluid fraction approaches unity. Here the speeds of sound (\ref{Eqcpm}) behavior has two cases:
 \begin{align}
 \left.\begin{array}{lr}
         c_+\to c_\kappa, & \\
          c_-\to c_{u}, & 
        \end{array}\right\}  & \qquad\mbox{for}\,\,\,c_\kappa>c_{u},\label{ckgtrcuu}\\
        \nonumber\\
 \left.\begin{array}{lr}
         c_+\to c_{u}, & \\
          c_-\to c_{\kappa}, & 
        \end{array}\right\}  & \qquad\mbox{for}\,\,\,c_\kappa<c_{u}.\label{cklesscuu}\
 \end{align}
   For the dipolar case $c_\kappa>c_{u}$, while  the soft-core case instead has $c_\kappa<c_{u}$ [see Figs.~\ref{figDipsound}(d) and \ref{figSCsound}(d)]. 
  
 These results emphasize an important difference between the two models we consider. The crystalline phase of the dipolar system is dominated by the compressive energy (i.e.~$mc_\kappa^2>mc_{u}^2$), while the soft-core system is dominated by the lattice compressive energy.

\subsubsection{Deep in the modulated regime}
\begin{figure}[htbp]
	\centering
	\includegraphics[width=3.3in]{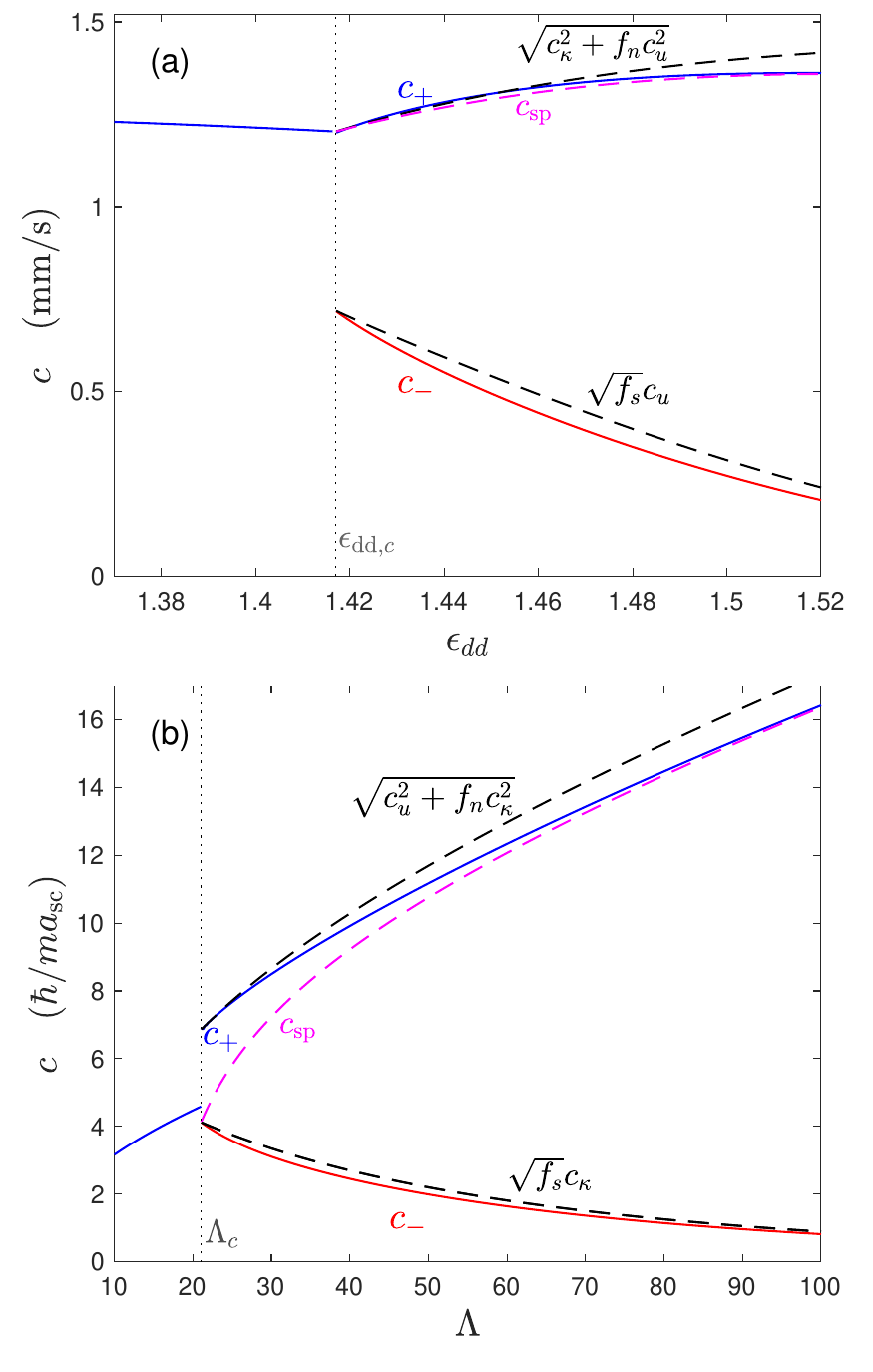}
	\caption{Comparison of limiting results for the speeds of sound for the (a) dipolar and (b) soft-core gases. BdG results (solid lines)  is compared to the limiting results (labelled).  Other parameters as in Figs.~\ref{figDipsound} and \ref{figSCsound}.
	\label{figlimits}}
\end{figure}

 Deep in the modulated phase the density contrast approaches unity and the superfluid fraction vanishes. This is known as the isolated droplet  or classical crystal regime, because the ability for atoms to tunnel between unit cells vanishes.
In this limit we have that  
\begin{align}
c_+&\to c_{\mathrm{sp}},\\
c_-&\to 0,
\end{align}
where we have introduced the characteristic speed of sound
\begin{align}
mc_{\mathrm{sp}}^2=\rho\alpha_{\rho\rho}-2\alpha_{\rho u}+\frac{\alpha_{uu}}{\rho},\label{mcsp2}
\end{align}
which we discuss further in Sec.~\ref{Sec:spring}. The results in  Figs.~\ref{figlimits}(a) and (b)  show that this estimate works well as the superfluid fraction vanishes, although it provides a reasonable estimate of $c_+$ even close to the transition for the dipolar case.

\subsubsection{Rigid lattice limit}
When the lattice compressibility is much larger than the bulk modulus, i.e.~$mc_{u}^2\gg mc_\kappa^2$, the lattice dynamics are heavily suppressed.  Neglecting the  $\alpha_{\rho u}$ term, we have
\begin{align}
c_- & \to \sqrt{f_s}c_\kappa,\\
c_+ & \to \sqrt{c_u^2+f_nc_\kappa^2},
\end{align}
where $f_n=1-f_s$. This limit is approximately applicable to the soft-core gas and the results for $c_\pm$ are shown in Fig.~\ref{figlimits}(b).

\subsubsection{Bulk incompressible limit}
On the other hand for $mc_{u}^2\ll mc_\kappa^2$, we have
\begin{align}
c_- & \to \sqrt{f_s}c_u,\\
c_+ & \to \sqrt{c_\kappa^2+f_nc_u^2}.
\end{align}
 This limit is approximately  applicable to the dipolar gas and the results for $c_\pm$ are shown in Fig.~\ref{figlimits}(a).

\subsection{Relationship to linear-chain model}\label{Sec:spring}
\begin{figure}[htbp]
	\centering
	\includegraphics[width=3.3in]{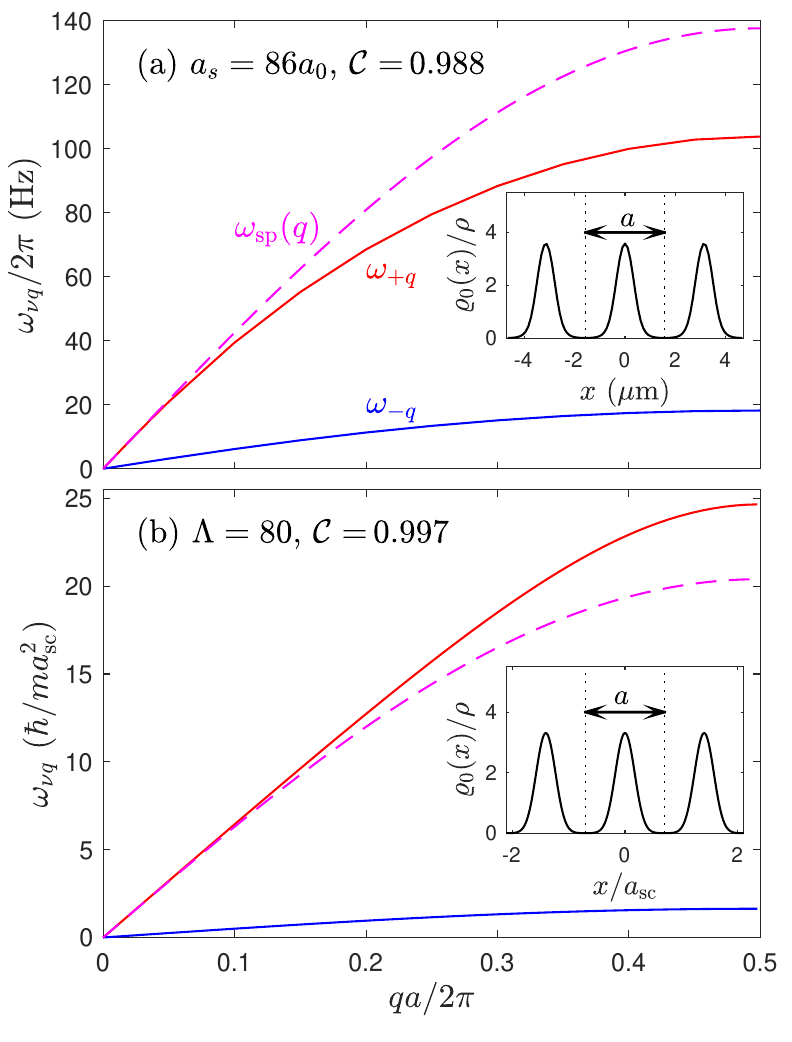}
	\caption{Comparison of the spring dispersion relation (dashed line) to the BdG results (solid lines) for (a) the dipolar model with $a_{s}=86\,a_0$ and (b) soft-core model with $\Lambda=80$. The insets show the linear density of the ground states for reference. Other parameters as in Figs.~\ref{figDipsound} and \ref{figSCsound}.
	\label{figspring}}
\end{figure} 

For a system deep in the modulated regime the ground state consists of atoms relatively localized in each unit cell (see insets to Fig.~\ref{figspring}). This regime should be well-approximated by the linear-chain model, in which a 1D solid is approximated by a set of masses joined by springs, yielding a simple theory for the phonons \cite{Kittel2004}. This idea was recently explored and found to be a good description of the dynamics of few droplets of a dipolar BEC confined by a three-dimensional cigar-shaped harmonic trap \cite{Mukherjee2023a}. 

 To apply this model to the systems we consider here, we denote the center-of mass position of the droplet at the $j$-th site by the coordinate $x_j$, with the equilibrium position being $x_j^0=ja$. In the linear-chain model the equation of motion for the droplet positions is 
\begin{align}
mN_\mathrm{uc} \ddot{x}_j=-k_\mathrm{sp}(2x_j-x_{j-1}-x_{j+1}),\label{Eq:spEoM}
\end{align}
where   $N_\mathrm{uc}=\rho a$ is the number of atoms  in each unit cell (localized droplet) and $k_\mathrm{sp}$ is the spring constant.  We identify the spring constant from how the energy of the droplet\footnote{Taking the droplet energy to be the energy in the unit cell, i.e.~$E=\mathcal{E}a$.}  $E$ varies with the inter droplet distance $a$:
\begin{align}
k_\mathrm{sp}=\left(\frac{\partial^2E}{\partial a^2}\right)_{N_\mathrm{uc}},
\end{align}
notably with $N_\mathrm{uc}$ held constant. This linear-chain model has a well-known phonon dispersion relation of  the form
\begin{align}
\omega_{\mathrm{sp}}(q)= \frac{2c_{\mathrm{sp}}}{a}\left|\sin\left(\frac{qa}{2}\right)\right| ,\label{linchaindisp}
\end{align}
where 
$
c_{\mathrm{sp}}=a\sqrt{ {k_{\mathrm{sp}}}/{mN_\mathrm{uc}}}$
is the speed of sound and $k$ is the wavevector. Evaluating this result in terms of the generalized elastic parameters yields Eq.~(\ref{mcsp2}) for $c_{\mathrm{sp}}$.

In Fig.~\ref{figspring} we compare the dispersion relations for the lowest two gapless bands obtained from the BdG calculations to the linear-chain dispersion relation (\ref{linchaindisp}), noting that the latter only depends on the elastic parameters $k_{\mathrm{sp}}$ obtained from  ground state solutions. The cases considered for these results are chosen to be deep in the crystal regime with $f_s\approx8\%$ (3\%) for the dipolar (soft-core) case. 
The results show that the low $q$ behavior of the upper gapless band is reasonably well predicted by $\omega_{\mathrm{sp}}(q)$ [also see~$c_{\mathrm{s}}$ and $c_+$ in Figs.~\ref{figDipsound}(b) and \ref{figSCsound}(b)]. At higher $q$ values, approaching the band edge,  the linear chain dispersion slightly overestimates (underestimates) the frequency of the upper excitation band  for dipolar (soft-core) model. These results confirm that the excitations in this band are closely related to the classical crystal phonon modes. 
  
\section{Density fluctuations}\label{Sec:DenFlucts}

\begin{figure*}[htbp]
	\centering
	\includegraphics[width=7in]{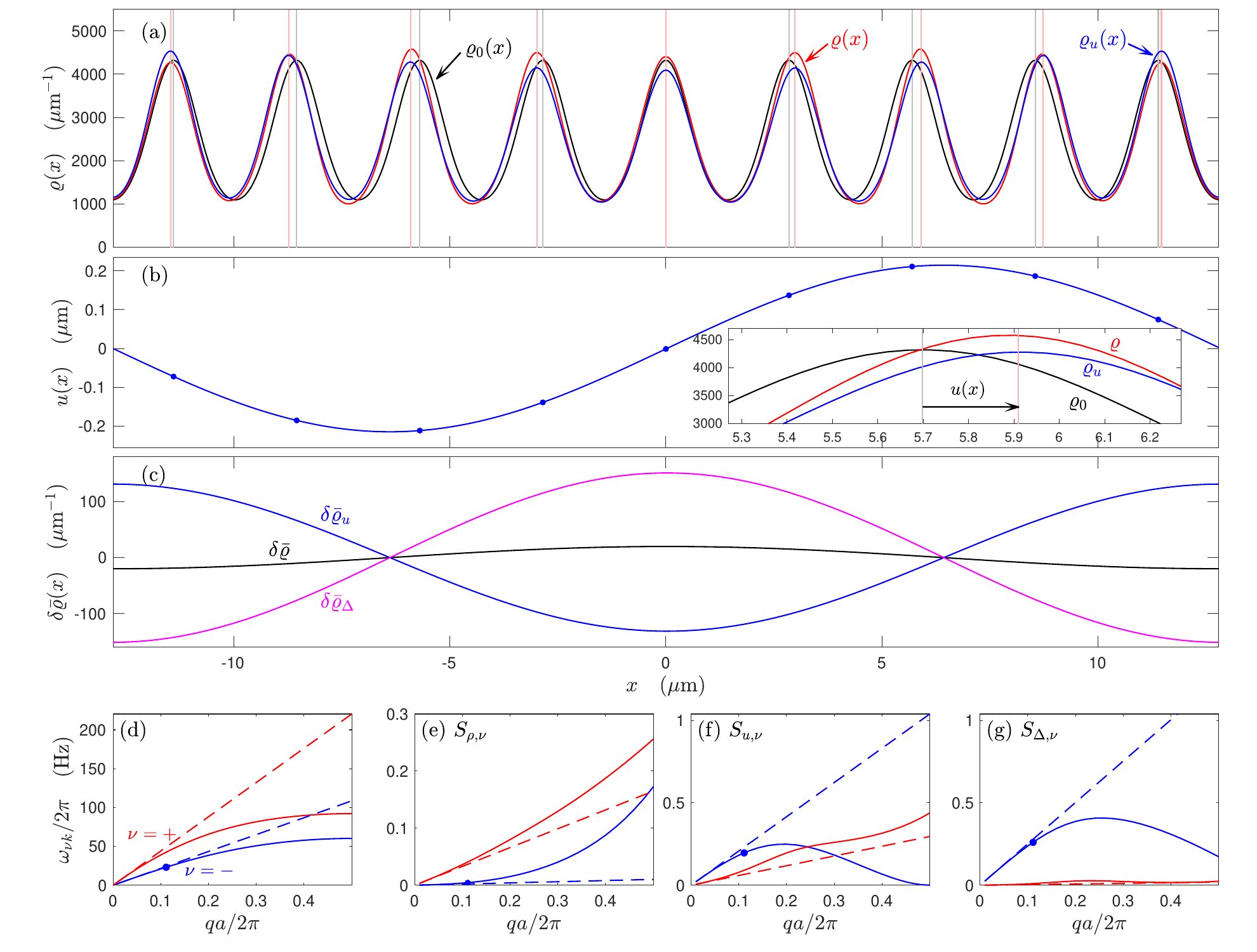}
	\caption{Effect of an excitation on the density fluctuations of a dipolar supersolid with $\epsilon_{\mathrm{dd}}=1.43$. (a) The ground state density $\varrho_0(x)$ (black) and density $\varrho(x)$  (red) when perturbed by a $\nu=1$ excitation with $q=2\pi/9a$ and $c_{\nu q}=5$. The  purely-deformed density profile $\varrho_u(x)$ (blue) using the deformation field determined from the quasiparticle. Vertical lines indicate the lattice locations for the ground state ($x_j^0$, grey lines) and the perturbed state ($x_j$, light red lines). (b) The deformation field $u(x)$ associated with the perturbed state from (a). Small circle markers indicate the deformation of the density maxima of $\varrho(x)$. Inset: close up of a density peak showing the identification of $u(x)$. (c) A comparison of the total, lattice deformation, and defect density fluctuations projected to the first Brillouin zone (see Appendix \ref{AppProjDensity}).  
	(d) Excitations frequencies for the lowest two bands comparing BdG (solid) and hydrodynamic [i.e.~$\omega_{\nu q}=c_\nu q$] (dashed) results. Marker identifies the excitation used in subplots (a)-(c). (e)-(f) The density fluctuations comparing the BdG (solid) and hydrodynamic (dashed) results.
	\label{figdenflucts}}
\end{figure*} 

We are interested in the density fluctuations caused by the quasiparticles of the gapless bands.
To do this we examine the effect of  adding an excitation with band index $\nu$ and quasimomentum $q$  to the condensate with a small coherent amplitude $c_{\nu q}$, i.e.
\begin{align}
\psi(\mathbf{x})  =  \psi_0(\mathbf{x})  +   c_{\nu q}\mathrm{u}_{\nu q}(\mathbf{x})   -c_{\nu q}^*\mathrm{v}^*_{\nu q}(\mathbf{x}).\label{psi_excite}
\end{align} 
In particular, we will consider the changes in the linear density
\begin{align}
\varrho(x)&=\int dy\,dz\,|\psi(\mathbf{x})|^2,
\end{align}
relative to the unperturbed case $\varrho_0(x)$.
Below we introduce the various density fluctuation measures and then consider the hydrodynamic limit of these results.

\subsection{Total density fluctuation}
The total density fluctuation  
\begin{align} 
\!\!\delta \varrho (x)&= 2\mathrm{Re}\left\{c_{\nu q}\!\int dy\,dz\,\psi_0(\mathbf{x})[\mathrm{u}_{\nu q}(\mathbf{x})-\mathrm{v}_{\nu q}(\mathbf{x})]\right\},
\label{deltavarho}
\end{align}
is the leading order expression in $c_{\nu q}$ for  $\varrho(x)-\varrho_0(x)$, and we have taken $\psi_0$ to be real.
In the modulated phase the density  varies rapidly with $x$, with a dominant periodicity set by the lattice constant $a$. 
Here we will focus on the density fluctuations arising on length scales much larger than the lattice constant, i.e.~with wave vectors in the first Brillouin zone. 
 For a single quasiparticle ($c_{\nu q}=1$) the only non-zero fluctuation is at wavevector $q$ and $-q$, with 
\begin{align}
\delta \tilde{\rho}_{\nu q} =\int d\mathbf{x}\,e^{-iqx}\psi_0(\mathbf{x})[\mathrm{u}_{\nu q}(\mathbf{x})-\mathrm{v}_{\nu q}(\mathbf{x})].\label{deltarhonuq}
\end{align} 
We show an example of  the   density change to a modulated ground state with the addition of a quasiparticle in  Fig.~\ref{figdenflucts}(a), and the associated long-wavelength total density fluctuation is visualized in Fig.~\ref{figdenflucts}(c) [also see Appendix \ref{AppProjDensity}].

\subsection{Lattice strain and associated density fluctuation}
With crystalline order we  can also consider the effect of the excitation (\ref{psi_excite}) on the crystal structure. This is characterized by the displacement field $u(x)$ which describes   the distance a point originally at $x$ displaces.  
We identify the lattice site locations as the local maxima of the linear density. The unperturbed locations being  $x_j^0=ja$ (i.e.~maxima of $\varrho_0$), and the locations for the perturbed case being $x_j=x_j^0+u(x_j^0)$  (i.e.~maxima of $\varrho$). The displacement field can be expressed as
\begin{align}
u(x)=2\mathrm{Re}\{c_{\nu q}u_{\nu q}e^{iqx}\}.\label{uxwave}
\end{align}
where the amplitude $u_{\nu q}$  is a measure of how the quasiparticle displaces the  site at the origin\footnote{In calculations we evaluate  $u_{\nu q}= -  \delta \rho_{\nu q}^\prime(0)/\varrho_0^{\prime\prime}(0)$, where the prime is derivative with respect to $x$ and  $\delta \rho_{\nu q}(x)=\int dy\,dz\,\psi_0 [\mathrm{u}_{\nu q} -\mathrm{v}_{\nu q} ]$}.  
In  Fig.~\ref{figdenflucts}(b) we show (\ref{uxwave}) for the perturbed density of Fig.~\ref{figdenflucts}(a), and give an example of its relationship to the site displacements [inset to Fig.~\ref{figdenflucts}(b)].

To understand the effect of lattice deformation on the density we consider a weak displacement field of the form (\ref{uxwave}) on a supersolid state with equilibrium density profile $\varrho_0(x)$. Assuming that the atom number in each cell remains constant then the purely-deformed density profile  is given by
\begin{align}
\varrho_{u}(x)= {\varrho_0\left(x-u(x)\right)}{\left(1-\partial_x u(x)\right)},\label{rhou}
\end{align}
where the first factor describes the displaced density profile and second factor adjusts the envelope of density profile to keep the atom number per cell constant.
The change in cell width, i.e.~strain, leads to a change in density across the system, giving rise to the lattice strain density fluctuation. Fourier transforming  $\delta\varrho_u(x)\equiv \varrho_u(x)-\varrho_0(x)$  for a single quasiparticle yields [cf.~Eq.~(\ref{deltarhonuq})] 
\begin{align}
\delta \tilde{\rho}_{u,\nu q}= -iq\rho u_{\nu q}L,\label{deltarhou}
\end{align} 
and $\delta \tilde{\rho}_{u,\nu -q}=\delta \tilde{\rho}_{u,\nu q}^*$ as the nonzero contributions in the  first Brillouin zone. In Eq.~(\ref{deltarhou})  $L$ is the length of the $x$ region integrated over.
 
We illustrate the lattice strain density fluctuation construction in Fig.~\ref{figdenflucts}. Using the identified deformation field $u(x)$ [Fig.~\ref{figdenflucts}(b)], in Fig.~\ref{figdenflucts}(a) we show the  purely-deformed density profile $\varrho_{u}(x)$, with associated  density fluctuation shown in  Fig.~\ref{figdenflucts}(c). 
 
\subsection{Defect density fluctuations}
In Fig.~\ref{figdenflucts}(a) we observe that while the maxima and minima of the density $\varrho(x)$ (i.e.~lattice distortion) coincide with those of $\varrho_{u}(x)$, these functions are not identical. The difference arises from the tunnelling of particles between sites, leading to a change in the atom number per site.   This is generally referred to as the  defect density \cite{Andreev1969a,Son2005a,Yoo2010a}, and can be defined as  $\varrho_{\Delta}(x)=\varrho(x)-\varrho_{u}(x)$, or equivalently
\begin{align}
\delta \tilde{\rho}_{\Delta,\nu q}=\delta\tilde{\varrho}_{\nu q}-\delta\tilde{\varrho}_{u,\nu q}.
\end{align}
We illustrate the defect density fluctuation in  Fig.~\ref{figdenflucts}(c).

\subsection{Density fluctuations of lowest bands}
 The decomposition of the total density fluctuation into strain and defect contributions is compared for an example case in Fig.~\ref{figdenflucts}(c). Here we see that the strain and defect densities have a relatively large amplitude and are out of phase, such that the total density fluctuation is  small. This behavior of the lattice straining in the opposite sense to the flow of atoms in the lowest gapless band reveals a general property of the lowest energy band we have used as an example in Fig.~\ref{figdenflucts}(a)-(c). Similar behavior has been observed in experiments with dipolar supersolids as a low frequency Nambu-Goldstone excitation   \cite{Tanzi2019b,Guo2019a,Natale2019a}.
 
It is useful to define the quantities
 \begin{align}
 S_{\rho,\nu}(q)&=\frac{|\delta \tilde\rho_{\nu q}|^2}{N},\label{Srho}\\
  S_{u,\nu}(q)&=\frac{|\delta \tilde\rho_{u,\nu q}|^2}{N},\\
  S_{\Delta,\nu}(q)&=\frac{|\delta \tilde\rho_{\Delta,\nu q}|^2}{N},\label{SDelta}
 \end{align}
to characterise the strength of the total density, strain density and defect density fluctuations, arising from the excitation with quantum numbers $\nu$ and $q$, where  $N=L\rho$. We note that $S_{\rho}(q)=\sum_{\nu} S_{\rho,\nu}(q)$ is the usual static structure factor, and
that aspects of the static and dynamical structure factors for the dipolar supersolid have been discussed in Refs.~\cite{Natale2019a,Petter2021a,Blakie2023a}.

In Fig.~\ref{figdenflucts}(c) we show the dispersion relations, and in Fig.~\ref{figdenflucts}(e)-(g) the related fluctuation strengths introduced in Eq.~(\ref{Srho})-(\ref{SDelta}), for the lowest two gapless bands of a dipolar supersolid. The lowest band ($\nu=-$, blue) has strong strain and defect density fluctuations as $q\to0$, but these are out-of-phase, leading to a small total density fluctuation. The second band ($\nu=+$, red) has weak defect density fluctuations, and the strain density fluctuation makes the dominant contribution to the total density fluctuation.  Hence the second band can be considered a crystal-like excitation, consistent with the conclusions from the linear-chain model presented in Sec.~\ref{Sec:spring}.

 \begin{figure}[htbp]
	\centering
	 \includegraphics[width=3.4in]{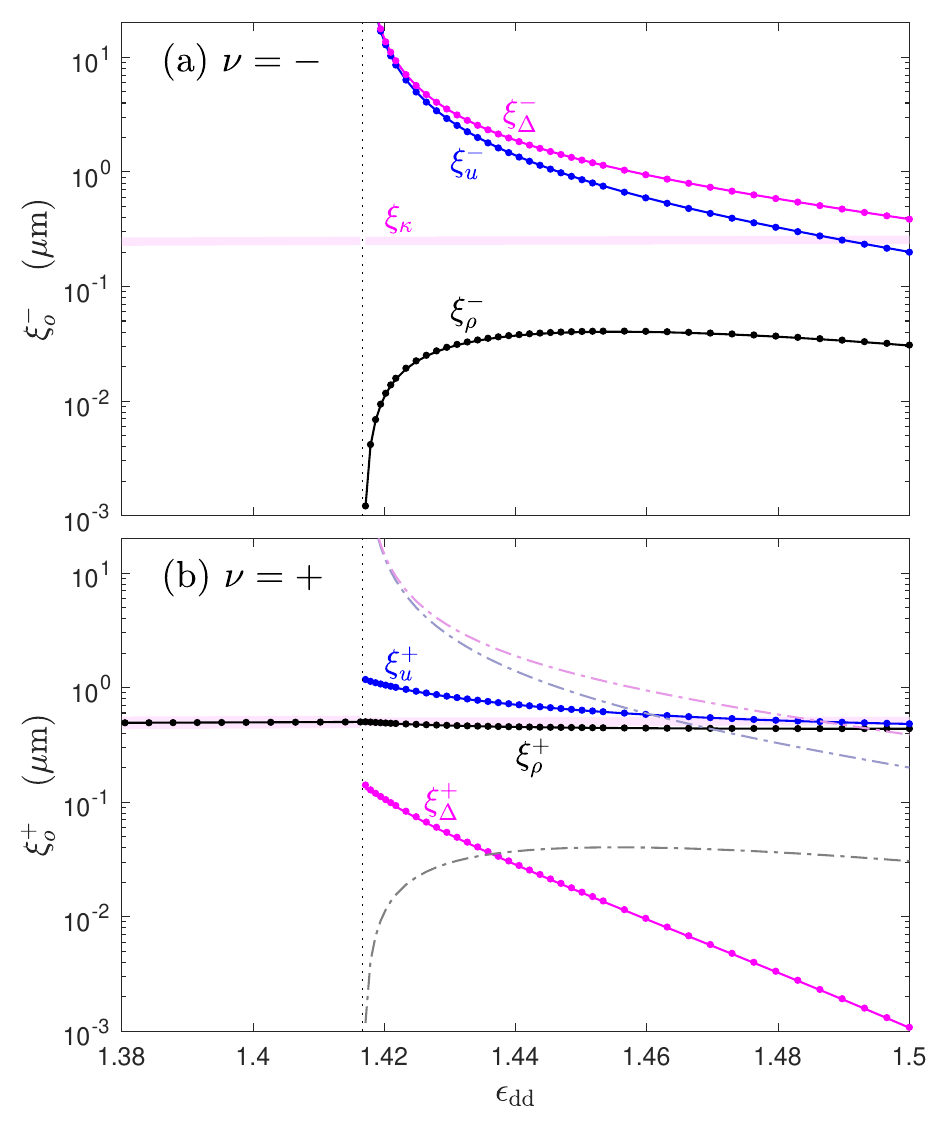} 
	\caption{Strengths of the hydrodynamic regime density fluctuations for a dipolar supersolid. (a) $\xi_o^+$ results (solid lines) compared to $\lim_{q\to0}2S_{o,+}(q)/q$ (markers) for the various density fluctuations. (b) $\xi_o^-$ results (solid lines) compared against $\lim_{q\to0}2S_{o,-}(q)/q$ (markers).  In both subplots we show $\xi_{\kappa}$ for reference. In (b) the results from (a) are repeated as light dash-dot lines. Other parameters as in Fig.~\ref{figDipsound}
	\label{figdipflucts}}
\end{figure}

\begin{figure}[htbp]
	\centering
	\includegraphics[width=3.4in]{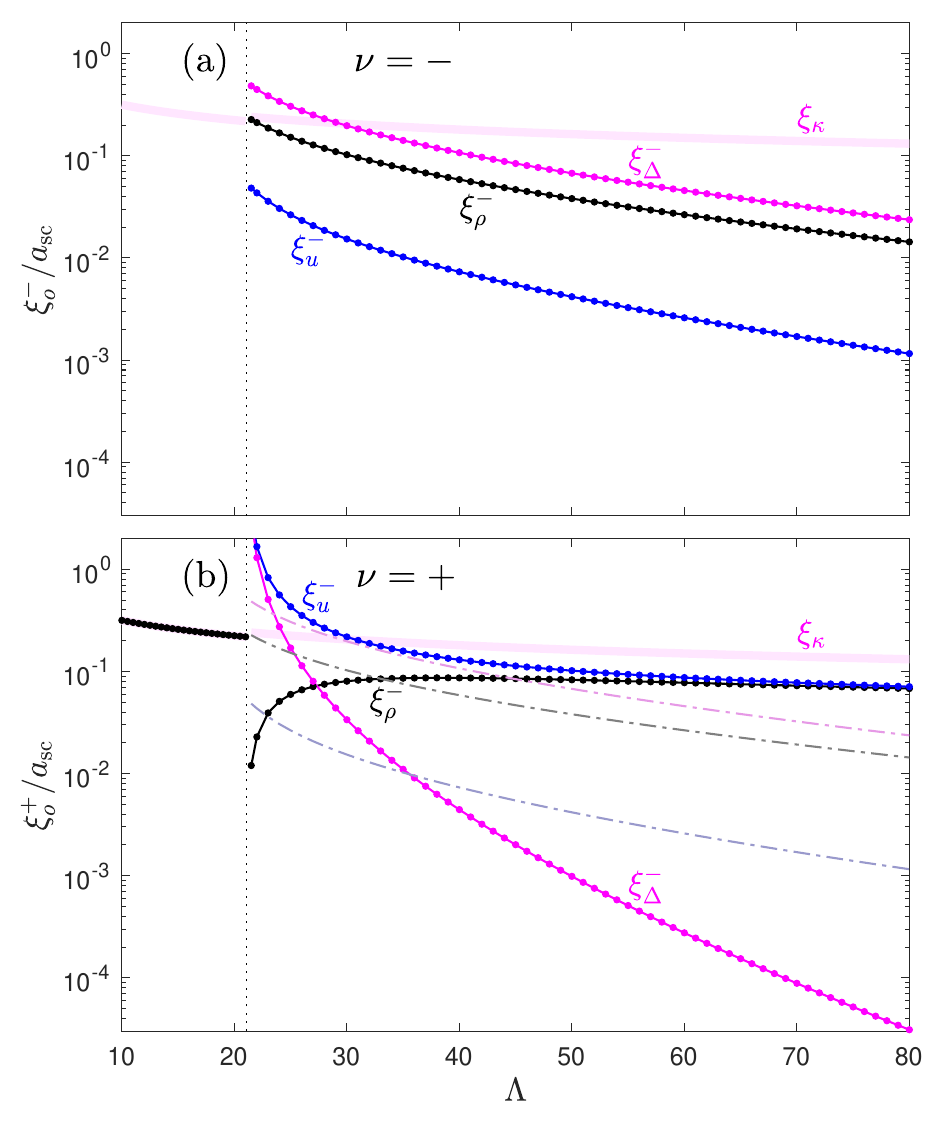}
	\caption{Strengths of the hydrodynamic regime density fluctuations for a soft-core supersolid. (a) $\xi_o^+$ results (solid lines) compared against $\lim_{q\to0}2S_{o,+}(q)/q$ (markers) for the various density fluctuations. (b) $\xi_o^-$ results (solid lines) compared against $\lim_{q\to0}2S_{o,-}(q)/q$ (markers).  In both subplots we show $\xi_{\kappa}$ for reference. In (b) the results from (a) are repeated as light dash-dot lines. Other parameters as in Fig.~\ref{figSCsound}.
	\label{figSCflucts}}
\end{figure}

The hydrodynamic description (\ref{LagrangianDensity}) also allows the calculation of the density correlation functions (see \cite{Yoo2010a}). In particular, our interest is in the imaginary part of the density response function, which has the form
\begin{align}
    \chi_{o}''( q,\omega) &= \frac{1}{2}Nq\pi\sum_\nu  \xi^\nu_o[\delta(\omega-c_\nu q)-\delta(\omega+c_\nu q)],\label{chiopp}
 \end{align}
 where the subscript $o$ can take the values $\{\rho,u,\Delta\}$ to represent the total density, strain density and defect density, respectively. This result is limited to the lowest two bands (i.e.~$\nu=\{-,+\}$) in the modulated regime and to a single band (i.e.~$\nu=\{+\}$) in the uniform regime. The hydrodynamic  results for the speeds of sound $c_\nu$  have already been introduced, and $\xi_o^\nu$ is a correlations length that represents the contribution of band $\nu$ to the response function. For the modulated case these are given by
 \begin{align}
  \xi_\rho^{\pm} &=  \frac\hbar{m\rho}\frac{\rho m c_\pm^2-\frac{\rho_s}{\rho_n}\alpha_{uu}}{mc_\pm(c_\pm^2-c_\mp^2)},\\
    \xi_{u}^{\pm} &=  \frac{\hbar\rho}{m\rho_n}\frac{ m c_\pm^2-\rho_s\alpha_{\rho\rho}}{mc_\pm(c_\pm^2-c_\mp^2)},\\
  \xi_{\Delta}^{\pm} &=  \frac{\hbar\rho_s}{m\rho\rho_n}\frac{\rho m c_\pm^2-(\alpha_{uu} - 2\rho\alpha_{\rho u} + \rho^2\alpha_{\rho\rho})}{mc_\pm(c_\pm^2-c_\mp^2)}.
\end{align}
In the uniform case the only non-zero quantity is $\xi_\rho^+\to\xi_\kappa$, with $\xi_\kappa=\hbar/mc_\kappa$ being the usual expression for the healing length of a Bose-Einstein condensate. 

At zero temperature  and  for $\omega>0$, the response function relates to the dynamic structure factor as $\chi_{o}''( q,\omega) =\pi S_{o}( q,\omega)$  (e.g.~see \cite{BECbook}), and from this we can obtain the static structure factors  $NS_{o}(q) =  \int d\omega S_{o}( q,\omega)$. 
The static structure factors can be decomposed as $S_{o}(q) = \sum_{\nu}  S_{o,\nu}(q)$
where $ S_{o,\nu}(q)$ is the contribution from band $\nu$, which were defined in  Eqs.~(\ref{Srho})-(\ref{SDelta}).
From hydrodynamic result  (\ref{chiopp}) we obtain
\begin{align}
\lim_{q\to0} S_{o,\nu}(q)=\frac{1}{\pi N}\int _0^\infty d\omega \,   \chi_{o}''( q,\omega)=\frac{1}{2}q\xi^\nu_o.\label{SonuHydro}
\end{align} 
 
In Figs.~\ref{figdenflucts}(e)-(g) we  compare the $S_{o,\nu}(q)$ functions obtained from the BdG calculations Eq.~(\ref{Srho})-(\ref{SDelta}) to the hydrodynamic result for a dipolar supersolid. All the density fluctuations vanish proportional to $q$  in the $q\to0$ limit, and the hydrodynamic result is seen to agree with the BdG results in this limit.
A survey over a broader parameter regime is shown in Figs.~\ref{figdipflucts} and \ref{figSCflucts} for the dipolar and soft-core systems, respectively. In these plots we only consider the  hydrodynamic character of the fluctuations and compare $\xi_o^\nu$   to the $q\to0$ limit of  $2S_{o,\nu}(q)/q$ calculated from the excitations\footnote{We use a BdG excitation with $q\sim10^{-2}/a$ to extract the limiting behavior.}.  

Our results show that strain and defect density fluctuations are reasonably strong in the $\nu=-$ band of the dipolar supersolid, but these contributions cancel out to suppress the total density functions [Fig.~\ref{figdipflucts}(a)]. In contrast for the same band of the soft-core supersolid  the lattice strain density fluctuations are heavily suppressed [Fig.~\ref{figSCflucts}(a)]. As noted in Sec.~\ref{Sec:Excitations},  these two systems differ in the relative size of their characteristic energies $mc_\kappa^2$ and $mc_{u}^2$. For the dipolar system the bulk compressibility dominates and it is favorable to reduce the total density fluctuations. In contrast for the soft-core system the lattice compressibility dominates and it is favourable to reduce lattice motion. Interestingly $\nu=+$ band dominates the lower band for total density fluctuations in the dipolar supersolid, whereas for the soft-core case the $\nu=-$ band dominates close to the transition.

For the   $\nu=+$ band the role of defect fluctuations in the modulated state is more important for the soft-core system close to the transition. However, for both models deep into the crystalline regime, the defect contribution is negligible and we find $\xi_u^+\approx\xi_\rho^+$. This is consistent with this band becoming a purely crystal excitation.

Finally, we note that the response functions relate to important sum rules. Notably, the 
 compressibility sum rule
 \begin{align}
 \int_{-\infty}^\infty\frac{d\omega}{\pi}\frac{\chi_{\rho}''( q,\omega)}{\omega}=\frac{N\alpha_{uu}}{\rho(\alpha_{\rho\rho}\alpha_{uu}-\alpha_{\rho u}^2)}=N\hbar\rho\kappa,
 \end{align}
 [cf.~Eq.(\ref{kappa})] and the
 $f$-sum rule
 \begin{align}
 \int_{-\infty}^\infty\frac{d\omega}{\pi}\omega\chi_{\rho}''( q,\omega)=  N\frac{\hbar q^2}{m}.
 \end{align}

\begin{figure}[htbp]
	\centering
	\includegraphics[width=3.4in]{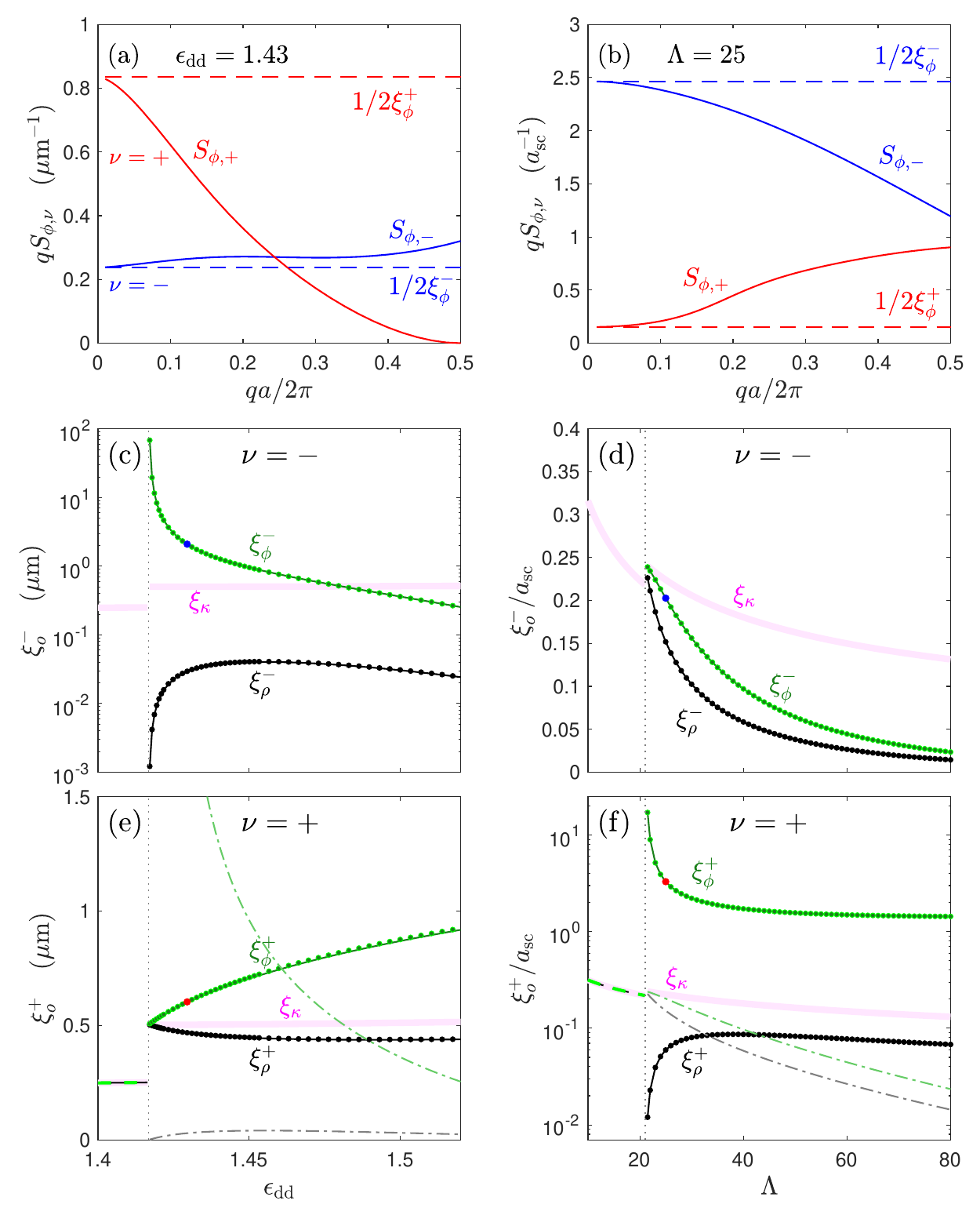}
	\caption{Phase fluctuations. (a), (b) Comparison of the phase fluctuations as a function of $q$ obtained from the BdG calculations (solid) to the hydrodynamic limit $\xi_\phi^\pm$ (dashed) for the lowest two bands (a) dipolar and (b) soft-core supersolids. (c)-(f) Strengths of the hydrodynamic regime total density $\xi_\rho^\pm$ (black line) and phase $\xi_\phi^\pm$  (green line) correlation lengths for a dipolar supersolid, compared to the relevant limits $\lim_{q\to0}1/[2qS_{\phi,\nu}(q)]$ (markers). Results for  the (c)  $\nu=-1$ and (e) $\nu=+1$ bands of a dipolar supersolid, with other parameters as in Fig.~\ref{figDipsound}.  In both subplots we show $\xi_\kappa$ for reference. Results for  (d)  $\nu=-1$ and (f) $\nu=+1$ bands of a soft-core supersolid, with other parameters as in Fig.~\ref{figSCsound}. In (e) and (f) the $\xi_\phi^+$ and $\xi_\rho^+$ are identical for the uniform phase, and are indicated by dashed lines. Also, the modulated results from (c) and (d) are repeated as light dash-dot lines.
	\label{figphflucts}}
\end{figure}

\section{Phase fluctuations}\label{Sec:PhaseFlucts}
We can also consider the effect of a quasiparticle excitation on the phase fluctuations.  
The phase is undefined as the density vanishes, so we avoid integrating over transverse coordinates and only consider the phase on the $x$ axis (i.e.~hereon we take $y=z=0$ in the dipolar model). For a quasiparticle perturbation of the form (\ref{psi_excite}) that phase fluctuation is $\phi(x)=\arg\left\{\psi(x)\right\}$ (since $\psi_0$ is taken to be real).
To leading order in $c_{\nu q}$ this is 
\begin{align}
    \phi(x)= \frac{\mathrm{Im}\{c_{\nu q}[\mathrm{u}_{\nu q}(x)+\mathrm{v}_{\nu q}(x)]\}}{\psi_0(x)}
\end{align}
For a single quasiparticle, the only non-zero fluctuations are at wavevectors $q$ and $-q$, with [cf.~Eq.~(\ref{deltarhonuq})] 
\begin{align}
 \tilde{ \phi}_{\nu q}= \int dx\,e^{-iqx}\frac{\mathrm{u}_{\nu q}(x)+\mathrm{v}_{\nu q}(x)}{2i\psi_0(x)}.
\end{align} 
The strength of the phase fluctuations are given by the dimensionless quantity
 \begin{align}
S_{\phi,\nu}(q)=\frac{\rho}{L}|\tilde\phi_{\nu q}|^2.\label{Sphi}
\end{align}

Similar to the treatment of density functions, from the quadratic Lagrangian we obtain the phase response function 
\begin{align}
    \chi_{\phi}''( q,\omega) &= \frac{1}{2}N \pi\sum_\nu  \frac{1}{\xi^\nu_\phi q}[\delta(\omega-c_\nu q)-\delta(\omega+c_\nu q)],\label{chiphipp}
 \end{align}
where  we have introduced the phase correlation length
\begin{align}
\xi_\phi^{\pm}  &=   \frac{\hbar}{\rho}
\frac{mc_\pm(c_\pm^2-c_\mp^2)}{\alpha_{\rho\rho} m c_\pm^2-\frac{1}{\rho_n}(\alpha_{\rho\rho}\alpha_{uu}-\alpha_{\rho u}^2)}.
\end{align}
This shows that the phase response diverges as $q\to0$. From this result we can derive the hydrodynamic result for the phase fluctuations 
\begin{align}
\lim_{q\to0}S_{\phi,\nu}(q)= \frac{1}{2\xi_\phi^{\pm}q}.
\end{align}

In Figs.~\ref{figphflucts}(a) and (b) we show examples of $S_{\phi,\nu}(q)$ computed from the BdG results for the dipolar and soft-core supersolid states, respectively. We also confirm that the hydrodynamic results agree in the $q\to0$ limit. In Figs.~\ref{figphflucts}(c)-(f) we examine the hydrodynamic limiting behavior of the fluctuations [i.e.~comparing $\xi_\phi^\nu$ to $\lim_{q\to0}q/2S_{\phi,\nu}(q)$] crossing the transition for both models.
In the uniform superfluid state $\xi_\phi^+\to\xi_\kappa$, being identical to the length scale for the total density fluctuations in the same regime. In the dipolar supersolid the $\nu=-$ band exhibits stronger phase fluctuations close to the transition, but as $\epsilon_{\mathrm{dd}}$ increases the phase fluctuations of the  $\nu=+$ band eventually dominates.  For the soft-core supersolid the $\nu=+$ band dominates the phase fluctuations everywhere. 

We note that the Fourier component of the superfluid velocity field is revealed from the phase fluctuations as $v_{\nu q}=i\hbar q \phi_q/m$, and thus our results are trivially extendible to describe the superfluid velocity fluctuations (also see \cite{BECbook}). The character of quasiparticle effect on the  phase or velocity has been employed to characterise the bands and excitations of dipolar supersolids (e.g.~see \cite{Natale2019a,Kirkby2023a}). However, these characterizations involve quantifying the phase variations distinguishing between the high and low density regions of the supersolid within the unit cell, and are thus beyond the hydrodynamic description.

\section{Outlook and Conclusions}\label{Sec:Conclusions}
 
In this study, we have demonstrated the effectiveness of the supersolid Lagrangian formalism in providing a quantitative description of the hydrodynamic characteristics of dipolar and soft-core supersolids. The Lagrangian description depends on a number of elastic parameters and we have discussed how these can be extracted from ground state calculations. Our findings include a comparative analysis of the Lagrangian theory to the BdG excitations for the speeds of sound, density and phase fluctuations. The excellent agreement shows that the hydrodynamic behaviour is well-described by the Lagrangian. Also, these results underscore the distinct behaviors exhibited by dipolar and soft-core models.

To unravel these distinctions, we have considered various limiting properties of the Lagrangian theory and how these apply to the two models. This has brought to light the significance of two energy scales, $mc_\kappa^2$ and $mc_{u}^2$, which delineate the relative impacts of bulk and lattice compressibility, respectively. The incompressible limit ($mc_\kappa^2\gg mc_{u}^2$) applies to the dipolar supersolid, while the rigid lattice limit ($mc_\kappa^2\ll mc_{u}^2$) applies to the soft-core supersolid. Our results illuminate how these distinct limits manifest in the excitation spectrum, density, and phase fluctuations.  

It is interesting to consider how our predictions can be verified in experiments.
Recently, the work of \v{S}indik \textit{et al.}~\cite{Sindik2023a} proposed a scheme for measuring the excitations and density response of a dipolar supersolid with minimal finite size effects. Their scheme involves the use of a toroidal potential for confinement with a well-defined mean density, and an azimuthally varying perturbing potential to excite $q\to0$ excitations. Observing the density oscillations following the sudden removal of the potential can be used to determine the speeds of sound and the amplitude of the density response. Determining the lattice strain and defect density fluctuations may be possible with high-resolution \textit{in situ} imaging which have already been used in experiments to characterize excitations \cite{Guo2019a} (also see \cite{Hertkorn2021a,Schmidt2021a}).  
  
Our results describe the zero temperature fluctuations and an important extension of this work is to include finite temperature effects (see Refs.~\cite{Yoo2010a,Hofmann2021a}). Another area of interest is higher dimensional supersolids where an elastic tensor describes the crystal, and the shear modulus will become relevant. Of interest is the two-dimensional supersolid that has recently been produced in experiments with a dipolar BEC \cite{Norcia2021a}. In the thermodynamic limit the ground state phase diagram for this case has been determined \cite{Zhang2019a,Ripley2023a}, yet the excitations remain largely unexplored. This system is expected to have three gapless branches \cite{Watanabe2012a} (cf.~\cite{Macri2013a}) with the emergence of a transverse crystal mode.

\emph{Note added.} As we were about to submit, the preprint Ref.~\cite{Rakic2024a} appeared, which also considers the elastic properties of softcore supersolids.

\section*{Acknowledgments}
\noindent PBB  acknowledges useful discussions with J.~Hofmann and  use of New Zealand eScience Infrastructure (NeSI) high performance computing facilities. LMP, DB and PBB, acknowledge support from the Marsden Fund of the Royal Society of New Zealand.
  
 \appendix 
\section{Additional details on supersolid models}\label{App:SSmodels}
Here we outline some additional details about the two models we consider and how we solve for the energy minimising states.

\subsection{Dipolar EGPE theory}\label{App:DEPGE}
The eGPE energy functional for this system is  
\begin{align}
E[\psi,a] &= \int_{\mathrm{uc}} d\bx\, \psi^*\left[H_\mathrm{sp}+\tfrac12\Phi(\mathbf{x})  +\tfrac25\gammaQF|\psi|^3\right]\psi,\label{Efunc}
\end{align}
where  \begin{align}
\Phi(\bx)=\int d\bx'\,U(\bx-\bx')|\psi(\bx')|^2,
\end{align}
and the effects of quantum fluctuations are described by the term with coefficient $\gammaQF = \frac{32}3 g_s\sqrt{a_s^3/\pi}\mathcal{Q}_5(\edd)$, with $\mathcal{Q}_5(x)=\Re\{\int_0^1 du[1+x(3u^2 - 1)]^{5/2}\}$ \cite{Lima2011a}.
The energy functional  is evaluated  in a single unit cell of length $a$ along $x$ (and integrated over all space in the transverse directions). The effective potential arising from interactions, $\Phi(\bx)$, is evaluated using a truncated interaction kernel  (see Ref.~\cite{Smith2023a} for details). 
The energy minimising wavefunction for (\ref{Efunc}) is determined using the gradient flow (or imaginary time) evolution \cite{Bao2004a,Lee2021a}, noting that the average density condition enforces the following normalization constrain on the wavefunction
\begin{align}
\int_{\mathrm{uc}}d\mathbf{x}\,|\psi(\mathbf{x})|^2=\rho a.
\end{align}
As a result we obtain $E(\rho,a)$, i.e.~the minimising energy as a function of density and cell size. We discuss the energy density function and identifying the ground state in Sec.~\ref{SecEdenGS}.
  
\subsection{1D soft-core formalism}\label{App:SC}
 The stationary states are determining by minimising the meanfield functional of the state
\begin{align}
{E}[{\psi},a]= \int_{\mathrm{uc}}dx\,\psi^*\left[-\frac{\hbar^2}{2m}\frac {d^2}{dx^2}+\frac{1}{2}\Phi_{\mathrm{sc}}(x)\right]\psi,\label{Esc}
\end{align}
where the integration is over a unit cell along $x$ of length $a$, and
\begin{align}
\Phi_{\mathrm{sc}}(x)=\int dx^\prime\,U_{\mathrm{sc}}(x-x^\prime)|\psi(x^\prime)|^2.
\end{align}
To implement the average density constraint on the stationary states $\psi$ has the normalization condition 
\begin{align}
\int_{\mathrm{uc}}  dx\,|\psi(x)|^2=\rho a.\label{scc}
\end{align}
Identical  to the dipolar case, we obtain the minimum energy per unit cell  $E(\rho,a)$, by applying gradient flow algorithm optimise the energy functional  Eq.~(\ref{Esc}) against the wavefunction.  
 
 \subsection{Energy density and ground state}\label{SecEdenGS}
 For both models are hence able to solve for energy minimising states with specified average linear density $\rho$ and lattice constant $a$.  The ground state $\psi_0$ for density $\rho$ is then determined by minimising this against $a$ [i.e.~Eq.~(\ref{E_0})].  
For determining the generalized elastic parameters we used the energy density (per unit length) we is given by
 \begin{align}
 \mathcal{E}(\rho,a)=\frac{{E}(\rho,a)}{a}.
 \end{align}
We only need to obtain $\mathcal{E}(\rho,a)$ for values of $a$ close to the ground state value at each density, since the elastic parameters involve derivatives evaluated at the ground state value.
We also note that in situations where the stationary state is uniform $\mathcal{E}$ independent of $a$. 

\section{Projected density fluctuations}\label{AppProjDensity}
 We reveal the long-wavelength character of the density fluctuations by projecting them to the first Brillouin zone with the operation
 \begin{align}
 \mathcal{P}_{\mathrm{BZ}}f(x)=\sum_{k\in\mathrm{BZ}}\frac{1}{L}\int dx^\prime e^{ik(x-x^\prime)}f(x^\prime),
 \end{align}
 where $\mathrm{BZ}$ denotes wavevectors in the range $[-\pi/a,\pi/a]$.
 For the unperturbed linear density only the  constant ($k=0$) term survives projection $\rho= \mathcal{P}_{\mathrm{BZ}}\varrho_0(x)$, i.e., the projected density is the (constant) mean linear density. We also define the projected total density fluctuation [shown in Fig.~\ref{figdenflucts}(c)]
 \begin{align}
 \delta\bar{\varrho}(x)= \mathcal{P}_{\mathrm{BZ}} \delta{\varrho}(x),
 \end{align}
 where for an excitation of quasimomentum $q$ only the $\pm q$ wavevectors contribute following projection, i.e.
 \begin{align}
  \delta\bar{\varrho}(x)=2\mathrm{Re}\{c_{\nu q}\delta \tilde{\rho}_{\nu q}e^{iqx}/L\},\label{delrarhoprojrel}
 \end{align}
with $\delta \tilde{\rho}_{\nu q}$ as defined in Eq.~(\ref{deltarhonuq}). We can similarly project the density fluctuations $ \delta {\varrho}_u(x)$ and  $\delta {\varrho}_\Delta(x)$ to obtain $ \delta\bar{\varrho}_u(x)$ and  $\delta\bar{\varrho}_\Delta(x)$, respectively  [also shown in Fig.~\ref{figdenflucts}(c)]. These satisfy similar relations to (\ref{delrarhoprojrel}) but in terms of $\delta \tilde{\rho}_{u,\nu q}$ and $\delta \tilde{\rho}_{\Delta,\nu q}$, respectively. For lattice strain density fluctuation this is
 \begin{align}
  \delta\bar{\varrho}_u(x)=2\mathrm{Re}\{-iq\rho c_{\nu q}u_{\nu q}e^{iqx}\},
 \end{align} 
 where we have used (\ref{deltarhou}). From this we see that $\delta\bar{\varrho}_u(x)=-\rho \,\partial_x u$ with $u$ given by Eq.~(\ref{uxwave}) \cite{Zippelius1980a}.

% \bibliography{dipolarpbb} 


\end{document}